\documentclass[aps,prl,twocolumn,showpacs]{revtex4}
\usepackage{graphicx}
\usepackage{dcolumn}
\usepackage{bm}
\begin{document}

{\bf Comment on \lq\lq Spectroscopic Evidence for Multiple Order Parameter
Components in the Heavy Fermion Superconductor CeCoIn$_5$\rq\rq}

Recently, Rourke {\it et al}. reported point-contact spectroscopy (PCS) results on the heavy-fermion superconductor CeCoIn$_5$ \cite{rourke05}. They obtained conductance spectra on the $c$-axis surfaces of CeCoIn$_5$ single crystals. Their major claims are two-fold: CeCoIn$_5$ has i) $d$-wave pairing symmetry and ii) two coexisting order parameter components. In this Comment, we show that these claims are not warranted by the data presented.

First, do their data represent spectroscopic properties of CeCoIn$_5$? Rourke {\it et al}. claim that their estimated contact {\it radius} satisfies the ballistic criterion \cite{sharvin65} at $T_c$ and even further at lower temperatures \cite{rourke05}. Our estimation using more rigorous formulas \cite{wexler66, blonder83deutscher94,pcsbasic} shows that their contact {\it diameter} ($d$), is larger than the mean free path ($l$) at $T_c$ by a factor of 1.2 -- 2.3, although $l/d \gg 1$ at lower temperatures. Since such an estimation (albeit a convention in the literature) just gives an indirect measure based on {\it bulk} parameters, it does not necessarily corroborate that a point contact formed on the {\it surface} is ballistic. The actual physical properties at the contact region can be much different from those in bulk, depending on the surface cleanness, roughness, contact pressure, etc. Therefore, whether atypical PCS data such as in Ref. \cite{rourke05} contain intrinsic spectroscopic information or not should be checked more carefully beyond such simple estimations.

Second, we point out that the zero-bias conductance peak (ZBCP) and subsequent dip-hump structure seen in Fig. 1(a) of Ref. \cite{rourke05}, which is the main feature they attribute to $d$-wave symmetry, has also been frequently observed in $s$-wave superconductors \cite{sheet05}. In Fig. 1(a), we show our own data obtained from epitaxial MgB$_2$ thin films. While other possibilities are open, including multiple contacts, a well-known origin for the dip structure near the gap edge is local heating due to the non-ballistic nature of the contact \cite{sheet05}.

Third, we have obtained PCS data from CeCoIn$_5$ single crystals along both (001) and (110) directions over wide temperature ranges \cite{park05prb,park05spie}. These data were taken reproducibly, well within the Sharvin limit, without showing any significant heating effects. They are consistent with each other and can be analyzed with a single order parameter. It is important to sample more than one crystallographic orientations to conclude the order parameter symmetry and if multiple order parameters exist.

\begin{figure}[t]
\includegraphics[scale=1.02]{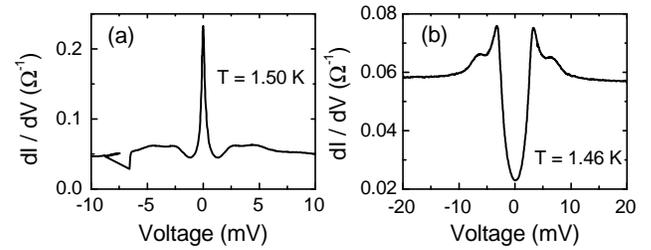}
\caption{\label{fig:mgb2PCS} Conductance spectra of point contacts on MgB$_2$ thin films using Au tips. (a) Reproducing the features in Refs. \cite{rourke05,sheet05}. (b) Reproducible data taken in the ballistic limit.}
\end{figure}

Finally, Rourke {\it et al}. base their claims of the $d$-wave symmetry on the ZBCP, which they attribute to Andreev bound states (ABS). It is well known that there are several origins for ZBCPs in tunneling conductance measurements \cite{greene00} and proper diagnostics must be performed to determine if a ZBCP actually arises from ABS, particularly tracking the evolution of the size and shape of the ZBCP with the magnitude and direction of an applied magnetic field \cite{greene00}. Measurements along different crystallographic orientations would also provide such information \cite{park05prb,park05spie}. Without such diagnostics, the origin of the ZBCP remains unknown. We also point out that other measurements classify CeCoIn$_5$ as either $d_{x^2-y^2}$ or $d_{xy}$, so ABS should not be observed on the $c$-axis surface of a single crystal. Rourke {\it et al}. compare their data with calculations using an extended Blonder-Tinkham-Klapwijk model, assuming parallel and serial combinations of conductance channels via surface ABS and bulk Andreev reflection. However, no materials micro-analysis is provided to justify their modeling and such a claim only supports the argument that the contacts are large and, thus, non-ballistic.

While we do not exclude the possibility of coexisting multiple order parameters in CeCoIn$_5$, we claim that Rourke {\it et al}.'s interpretation of their PCS results \cite{rourke05} as such evidence should be viewed critically.  \\

\noindent
W. K. Park and L. H. Greene \\
\indent Department of Physics and Frederick Seitz Materials
\indent Research Laboratory, University of Illinois at Urbana-
\indent Champaign, Urbana, Illinois 61801, USA \\

\noindent
Received 20 July 2005





\begin{references}

\bibitem{rourke05}
P. M. C. Rourke {\it et al}., Phys. Rev. Lett. {\bf 94}, 107005 (2005).

\bibitem{sharvin65}
Yu. V. Sharvin, Sov. Phys. JETP {\bf 21}, 655 (1965).

\bibitem{pcsbasic}
Yu. G. Naidyuk and I. K. Yanson, J. Phys.: Condens. Matter {\bf 10}, 8905 (1998); Yu. G. Naidyuk and I. K. Yanson, {\it Point-Contact Spectroscopy} (Springer, New York, 2005).

\bibitem{wexler66}
G. Wexler, Proc. Phys. Soc. London {\bf 89}, 927 (1966).

\bibitem{blonder83deutscher94}
G. E. Blonder and M. Tinkham, Phys. Rev. B {\bf 27}, 112 (1983); G. Deutscher and P. Nozi{\`e}res, {\it ibid}. {\bf 50}, 13557 (1994).

\bibitem{sheet05}
G. Sheet and P. Raychaudhuri, cond-mat/0502632.

\bibitem{park05prb}
W. K. Park {\it et al}., Phys. Rev. B {\bf 72}, 052509 (2005).

\bibitem{park05spie}
W. K. Park {\it et al}., Proc. SPIE {\bf 5932}, 59321Q (2005). {\it cf}. cond-mat/0507353.

\bibitem{greene00}
L. H. Greene {\it et al}., Physica B {\bf 280}, 159 (2000); L. H. Greene {\it et al}., Physica C, {\bf 408-410}, 804 (2004).

\end{references}
\end{document}